# Distribution of scientific journals impact factor

Michael Romanovsky

*slon@kapella.gpi.ru*
MIREA - Russian Technological University, Vernadsky Avenue 78, 119454 Moscow  (Russia)

**Abstract**
We consider distributions of scientific journals impact factor. Analysing 9028 scientific journals with the largest impact factors, we found that the distribution of them is year-to-year stable (at least for analysed 2011-2013 years), and it has the character of the exponential Boltzmann distribution with the power law asymptotic (tail).

**Introduction**

The impact factor is one of the main characteristic of scientific journals. Generally, it describes the relevance of a journal with respect to current scientific interests. Let us remind that the impact factor of the journal is the ratio of citations during 2 last years to the number of articles during the same 2 last years at this journal.

The distribution of journals impact factors is interesting in several aspects. First of all, the presence or absence of the stability is the crucial quality of such distributions. Indeed, the absence of the year-to-year stability raises the question of applicability of impact factor for characterization of the scientific journal system (and the system of obtaining the scientific knowledge) at all. Note that the impact factor of a definite journal itself may strongly change year-to-year. Secondly, the shape of this distribution may help to establish some internal scaling. Thirdly, even the phenomenological description of distribution permits to clarify stable characteristics that may be attracted to explain the natural separation of journals.

We use for calculations the data from the website https://www.citefactor.org/journal-impact-factor-list-2015_18.html. It contains 9028 scientific journals totally with the largest impact factors. Analysed years were 2011, 2012, and 2013

**Empirical data**

The results of empirical data extraction from above website are presented on Fig.1. We exclude from consideration one journal: *Ca-A Cancer Journal For Clinicians* since it impact factor for 2011-2013 strongly exceeds (about twice, namely it was 101,78; 153,459; 162,5 respectively) the impact factors of all other journals. The Fig.1 presents the cumulative distributions (i.e. the quantity of journals with the impact factor larger than the definite value) for 2011-2013. It is seen that the coincidence of three empirical distributions is very high instead of the fact that the impact factor of some journal may vary in wide range up to several units and moreover tens units (like above *Ca-A Cancer Journal For Clinicians*). The tails of distributions corresponded to impact factors larger than 50 have vague character due to very small number of large impact factor journals.

The practical coincidence of distributions for 2011-2013 years means that the distribution of impact factor is year-to-year stable (at least quasi-stable). As it was noted in **Introduction**, this permits to say that once introduced description journal system: the journals dividing on quartiles or on another parts – may be used for long year period.

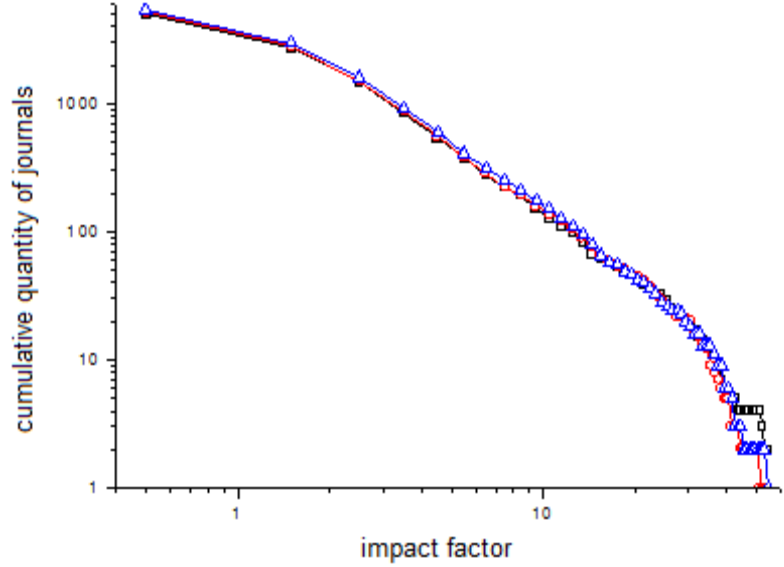

**Figure 1. The natural cumulative distributions of impact factor of scientific journals
(open triangles - 2013, open circles – 2012, open squares – 2011).**

It is seen also that the quantity of journals with impact factor less than one is larger than the half of total quantity, the number of journals with impact factor larger than ten is slightly more than one hundred.

**Phenomenological description**

Upon possibility, even the phenomenological description may provide more deep understanding of processes leading to one of the other of definite journal impact factor. Natural distributions on Fig.1 look like the exponential Boltzmann function with the power law asymptotic tail. The theory of such functions was done by (Garanina & Romanovsky, 2015(1)), there appeared also in the problem of citation of individual scientists (Garanina & Romanovsky, 2015(2)).

To define the general form of the desired distribution, one may proceed from the results presented in (Vidov & Romanovsky, 2011) as a starting point. The necessary transformation yields the curve with the exponential main part and a transition to power law at the tail in an explicit form of a probability distribution function (PDF):

$$W_{T\beta\theta}(R) = \frac{1}{\sqrt{\pi T}} \int_0^\infty \cos(x\sqrt{R}) \left\{ \frac{2}{\Gamma(\beta-1/2)} \left[ (\beta - 3/2) \frac{xT}{4\theta} \right]^{\beta/2 - 1/4} K_{\beta-1/2}\left[ \sqrt{(\beta - 3/2) \frac{xT}{4\theta}} \right] \right\}^\theta dx \quad (1)$$

Here $R$ is variable (the impact factor in our case, see below), $\Gamma$ is the gamma-function, $K_{\beta-1/2}$ is the modified Bessel function of the 2$^{nd}$ kind (also known as ''McDonald function''), $\theta$, $T$, $\beta$ are parameters. Note that the above integral converges for any $\beta>3/2$. It does not exist for smaller $\beta$. The fact that $W_{T\beta\theta}$ is positive for any $\beta>3/2$ is followed from the deduction of an initial probability density function (PDF) for $R$ in (Vidov & Romanovsky, 2011). It is clear that the expression for PDF like (1) can involve dependencies more complicated than the square root function. For the power-law function $W(R) \sim R^\xi$, ($\xi \neq 1/2, 1$), the final distribution (similar to that expressed by Eq. (1)) will have the stretch exponent main part with a power-law tail (Garanina & Romanovsky, 2015 (2)) Even more artificial forms of (1) may prove useful (see also (Romanovsky, 2009) for more on functional random walks). We will need also the cumulative distribution function $F_{T\beta\theta}(R) = \int_R^\infty W_{T\beta\theta}(R')dR'$.

The approximation of Eq. (1) for comparably small $R$ (up to several values of $T$) is easily reduced to a dependence on parameter $T$ only

$$W_T(R) \cong \frac{1}{T} \exp\left(-\frac{R}{T}\right) \quad (2)$$

As well, $F_T(R) \cong \exp(-\frac{R}{T})$ for small $R$. To obtain a general form of $W$, note that $K_{\beta-1/2}$ is defined through elementary functions for natural $\beta$. We need the function with $\beta=2$ only:

$$W_{T2\theta}(R) = \frac{1}{\sqrt{\pi T}} \int_0^\infty \cos(x\sqrt{R}) \exp\left(-x\sqrt{\frac{\theta T}{2}}\right)\left(1 + x\sqrt{\frac{T}{2\theta}}\right)^\theta dx \qquad (3)$$

with the corresponding function $F_{T2\theta}(R)$. Fig.2 demonstrates the comparison of natural distributions for 2011-2013 and the function $F_{T2\theta}(R)$:

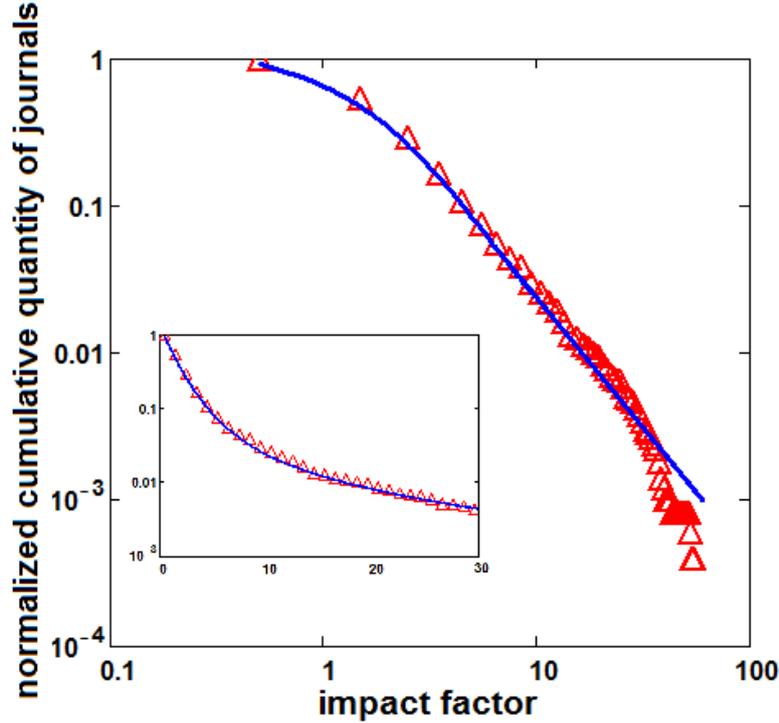

**Figure 2. The comparison of natural impact factor distributions 2011-2013 with the function $F_{T2\theta}(R)$. The main panel is drawn in double-logarithmic scale, the inset is the semi-logarithmic plot. The solid line is the function $F_{T2\theta}(R)$ with $T = 1,5$, $\theta = 30$, open triangles is the normalised cumulative distribution of impact factor for 2013.**

The good coincidence of empirical distribution with $F_{T2\theta}(R)$ shows that the natural distributions have the internal scaling parameter $T = 1,5$ for small values of impact factor, the distributions here are look like the exponential Boltzmann distribution $F_T(R)$. This parameter $T = 1,5$ may be used for natural separation of journals into some *strata*.

The parameter $\theta$ characterized the transition between exponential Botzmann part and power-law tail. The distribution Fig.2 looks like the known distributions of income and wealth (Yakovenko & Rosser, 2009), new car sells (Garanina & Romanovsky, 2015(1)), citations of individual scientist with moderate total number of citations (Garanina & Romanovsky, 2015(2)).

**Conclusion**.

The distribution of impact factor of scientific journal is year-to-year stable and may be characterised by two parameters: some "effective temperature" $T = 1,5$ that may be applied for small value of impact factor, and the parameter $\theta = 30$. The last parameter describes the transition between exponential part of distribution, and the asymptotic power-law part.

The presence of some effective temperature, or an internal scale in distribution may be used for construction of some separation system of scientific journals. The elevation of power law asymptotic tail over the pure exponent $F_T(R)$ contains about 100 journals only with the largest impact factors.

The impact factor as a random value belongs to the wide class of phenomena like sells, incomes, wealth, etc. that arise due to human activity (see also (Yakovenko & Rosser, 2009)).